\documentclass[prb,showpacs,twocolumn,amsmath,amssymb]{revtex4}
\usepackage{epsfig}

\begin{document}

\title{Multi-magnon bound states in the frustrated ferromagnetic 1D chain}

\author{Lars Kecke}
\affiliation{Condensed Matter Theory Laboratory, RIKEN, Wako,
Saitama 351-0198, Japan}
\author{Tsutomu Momoi}
\affiliation{Condensed Matter Theory Laboratory, RIKEN, Wako,
Saitama 351-0198, Japan}
\author{Akira Furusaki}
\affiliation{Condensed Matter Theory Laboratory, RIKEN, Wako,
Saitama 351-0198, Japan}

\date{August 3, 2007}

\begin{abstract}
We study a one-dimensional Heisenberg chain with competing ferromagnetic
nearest-neighbor and antiferromagnetic next-nearest neighbor interactions
in magnetic field.
Starting from the fully polarized high-field state, we calculate
the dispersions of the lowest-lying $n$-magnon excitations
and the saturation field ($n=2,3,4$).
We show that the lowest-lying excitations are always bound multi-magnon
states with a total momentum of $\pi$ except for a small parameter range.
We argue that bose condensation of the bound $n$ magnons
leads to novel Tomonaga-Luttinger liquids with multi-polar correlations;
nematic and triatic ordered liquids correspond to $n=2$ and $n=3$.
\end{abstract}
\maketitle

Frustrated spin chains are simple models that still show
surprisingly rich physics.
A prototype of these is the spin-$\frac12$ Heisenberg chain with
nearest-neighbor (NN) $J_1$ and next-nearest-neighbor (NNN) $J_2$
couplings,
\begin{equation}
H=\sum_{l\in\mathbb{Z}}\left(
J_1\vec{S}_l\cdot\vec{S}_{l+1}
+J_2\vec{S}_l\cdot\vec{S}_{l+2}
-hS^z_l\right),
\label{hamiltonian}
\end{equation}
where $\vec{S}_l=(S_l^x,S_l^y,S_l^z)$ is the spin-1/2 operator on
the $l$th site, and external field $h$ is applied in
the $z$ direction.
The exchange interactions are frustrated as
the NNN interaction is antiferromagnetic (AF), $J_2>0$.
Until recently most theoretical studies on the $J_1$-$J_2$ model
(\ref{hamiltonian})
considered the case where the NN coupling $J_1$ is also AF.
However, interest is now growing rapidly in
the ferromagnetic (FM) case ($J_1<0$) as well, which is
triggered by experimental reports on thermodynamic
properties of various quasi-one-dimensional frustrated FM spin chains
(for a list of frustrated quasi-one-dimensional materials, see
Table 1 in Ref.~\onlinecite{hase}
and Fig.~5 in Ref.~\onlinecite{drechsler}).
For example,
Rb$_2$Cu$_2$Mo$_3$O$_{12}$ (Ref.~\onlinecite{hase}) and LiCuVO$_4$
(Ref.~\onlinecite{enderle}) are considered to be described by the
$J_1$-$J_2$ model with $J_1\approx-3J_2$ and $J_1\approx-0.3J_2$,
respectively.

Recent theoretical studies\cite{vekua,kolezhuk} have shown that,
in a magnetic field, the FM $J_1$-$J_2$ chain ($J_1<0$, $J_2>0$) becomes
a Tomonaga-Luttinger (TL) liquid having nematic quasi-long-range order
as dominant correlation for some range of $J_1/J_2$.
This nematic state can be thought of as arising from condensation
of two-magnon bound states.\cite{chubukov}
Interestingly, such a nematic ordered phase can also appear in a frustrated
FM spin model on the two-dimensional square lattice.\cite{shannon}
One can imagine further a bose condensed phase of more-than-two-magnon
bound states. Indeed it was shown\cite{momoi} recently that
an octupolar-like triatic ordered phase can result from
condensation of three-magnon bound states in a frustrated ferromagnet
on the triangular lattice.
These results motivated us to examine the possibility of such many-magnon
bound states in the FM $J_1$-$J_2$ model (\ref{hamiltonian}).
In this paper we show, by explicitly constructing bound-state wave
functions, that for $-3.52\lesssim J_1/J_2\lesssim-2.72$
the lowest-lying excitations
from the fully polarized FM state are 3-magnon bound states,
and moreover 4-magnon bound states appear
for a slightly stronger FM coupling regime.
We then suggest exotic TL liquid phases with multipolar-like spin
correlations to emerge from condensation of multi-magnon bound states
below the saturation field.

Before presenting our calculations, let us briefly review
known results that are relevant to our study.
First, in the classical limit, we may regard $\vec{S}_l$ as
a c-number vector of length $S$.
The ground state of (\ref{hamiltonian}) in this limit
has a (right- or left-winding) helical spin structure,
$\vec{S}_l/S=(\sin\theta\cos\phi_l,\sin\theta\sin\phi_l,\cos\theta)$,
with a pitch angle of $\phi=\phi_{l+1}-\phi_l=
\pm\arccos\,(-J_1/4J_2)$
and a canting angle of
$\theta=\arccos\,[4hJ_2/S(J_1+4J_2)^2]$, so that
spins are fully polarized
at a saturation field of $h_s=S(J_1+4J_2)^2/4J_2$ if $-4J_2<J_1<0$,
or at zero field if $J_1<-4J_2$.
The helical order does not survive quantum fluctuations.

In the quantum spin-$\frac12$ case, the ground state is fully
polarized without magnetic field if $J_1<-4J_2$. At the boundary
$J_1=-4J_2$ the zero-field ground state is highly degenerate such
that states from vanishing magnetization
to full polarization share the same energy.\cite{hamada}
For the parameter range $-4J_2<J_1<0$ of our main interest,
Chubukov\cite{chubukov} suggested that the ground state just below
the saturation field should be a nematic state made up of bound magnon
pairs with a commensurate total momentum $k=\pi$ if
$-2.67J_2<J_1<0$ ($2.67\approx1/0.38$)
and with an incommensurate momentum $k<\pi$ otherwise,
which was partly verified by mean-field theory,\cite{dmitriev}
numerical study,\cite{honecker}
Green's function analysis which fixed the commensurate-incommensurate
transition point to $J_1/J_2=-2.66908 \,(=-1/0.374661)$,\cite{kuzian}
and weak-coupling bosonization analysis.\cite{vekua,honecker,cabra}
While earlier calculations of the ground-state magnetization process
suggested metamagnetic transitions,\cite{aligia,dmitriev}
recent density-matrix renormalization group (DMRG) study\cite{honecker}
finds that the total magnetization of finite-size chains changes
by $\Delta S^z=2$ at $J_1=-J_2$,  $\Delta S^z=3$ at $J_1=-3J_2$,
and $\Delta S^z=4$ at $J_1=-3.75J_2$
below saturation, implying that the magnetization curve is
continuous in the thermodynamic limit.\cite{cabra}

To reveal how the fully polarized FM state collapses into new states
with decreasing either magnetic field or the coupling ratio $|J_1|/J_2$,
we analyze magnon instability in the fully polarized state.
We apply a large enough magnetic field $h$
such that the fully polarized state is the unique ground state
for $-4J_2<J_1<0$,
and all multi-magnon excitations have positive excitation energies
which decrease as $h$ is reduced.
The saturation field $h_s$ is then defined as the field at which
the lowest excitation energy vanishes.
We consider bound states of
up to four magnons in a finite chain of $N$ spins.

To explain our computational scheme,
we begin with one- and two-magnon states.
The one-magnon excited state with momentum $k$,
\begin{equation}
\left|k\right>=
\frac{1}{\sqrt{N}}\sum^N_{l=1}e^{ikl}S_l^-\left|\mathrm{FM}\right>,
\end{equation}
on the fully polarized state $\left|\mathrm{FM}\right>=
\left|\uparrow\uparrow\uparrow\dots\uparrow\right>$
has the excitation energy
\begin{equation}
\epsilon_1(k)=J_1(\cos k-1)+J_2[\cos(2k)-1]+h,
\label{epsilon_1(k)}
\end{equation}
which has a minimum of
\begin{equation}
\epsilon_1(k_0)=-\frac{1}{8J_2}(J_1+4J_2)^2+h
\label{e1mag}
\end{equation}
at $\cos k_0=-J_1/4J_2$.
Obviously the one-magnon instability just
reproduces the classical helical spin state in the applied field
below the saturation field.
However, as shown by earlier
studies,\cite{chubukov,dmitriev,honecker,kuzian,vekua}
this is not the true instability for the quantum case
in the whole parameter region $-4J_2<J_1<0$.

To calculate bound $n$-magnon excitations, we take $n$-magnon
states with a center-of-mass (CM) momentum $k$ as a basis.
For example, our basis for the two-magnon excitations has
two $\downarrow$ spins with a total momentum $k$ and
a relative distance $r$
($=1,2,\ldots$),
\begin{equation}
\left|r;k\right>=\frac{1}{\sqrt{N}}\sum_le^{ik(2l+r)/2}S_l^-S_{l+r}^-
\left|\mathrm{FM}\right>\,.
\end{equation}
In this basis the matrix elements of the
Hamiltonian can be written as
$\langle r;k|H|r';k'\rangle=\delta_{k,k'}H_{r,r'}$,
where non-vanishing entries of $H_{r,r'}$ are
\begin{eqnarray}
&&
H_{r,r} =
J_1(\delta_{r,1}-2)+J_2(\delta_{r,1}\cos k+\delta_{r,2}-2)+2h,
\nonumber\\
&&
H_{r,r+1} = H_{r+1,r} = J_1\cos(k/2),
\label{H_r,r'}\\
&&
H_{r,r+2} = H_{r+2,r} = J_2\cos k.\nonumber
\end{eqnarray}
By separating off the CM motion we have reduced the two-magnon
problem to a one-particle one which in principle can be solved
exactly.
For general $k$ this involves finding roots of a
transcendental equation.\cite{kuzian}
The eigenvalue problem (\ref{H_r,r'}) is greatly simplified at $k=\pi$,
where the excitation energy
of the two-magnon bound state is\cite{chubukov,kuzian}
\begin{equation}
\epsilon_2(\pi)=-J_1-3J_2+\frac{J_2^2}{J_1-J_2}+2h.
\label{e2mag}
\end{equation}
To calculate the energy dispersion $\epsilon_2(k)$ of the two-magnon
bound states for general $k\in[0,\pi]$,
we numerically diagonalize the matrix $H_{r,r'}$ by restricting
$r$ and $r'$ up to $1000$.
As long as the maximum value of $r$ is sufficiently larger than the
size of bound states, finite-size corrections should be exponentially small.

The bound states of more than two magnons can be calculated
in a similar manner.
For the $n$-magnon sector, we take, as a basis set, states
with total momentum $k$ in which
$n$ down spins are separated by distance $r_1$, $\ldots$, $r_{n-1}$.
For example, the 3-magnon basis is given by
\begin{equation}
\left|r_1,r_2;k\right>=
\sum_l
\frac{e^{ik(3l+2r_1+r_2)/3}}{\sqrt N}
S_l^-S_{l+r_1}^-S_{l+r_1+r_2}^-\left|\mathrm{FM}\right>,
\end{equation}
with which the matrix elements
$\langle r_1,r_2;k|H|r_1',r_2';k\rangle$ are easily found.
The 4-magnon basis states $|r_1,r_2,r_3;k\rangle$
are constructed similarly.
To solve the 3- and 4-magnon bound states, we numerically diagonalized
the Hamiltonian matrix expressed in terms of the finite number of basis
states $|r_1,r_2;k\rangle$ with $1\le r_i\le27$
and $|r_1,r_2,r_3;k\rangle$ with $1\le r_i\le9$, respectively.
This seems to be sufficient to determine the lowest excitations.

\begin{figure*}
\begin{minipage}{8.5cm}
\centerline{\epsfig{file=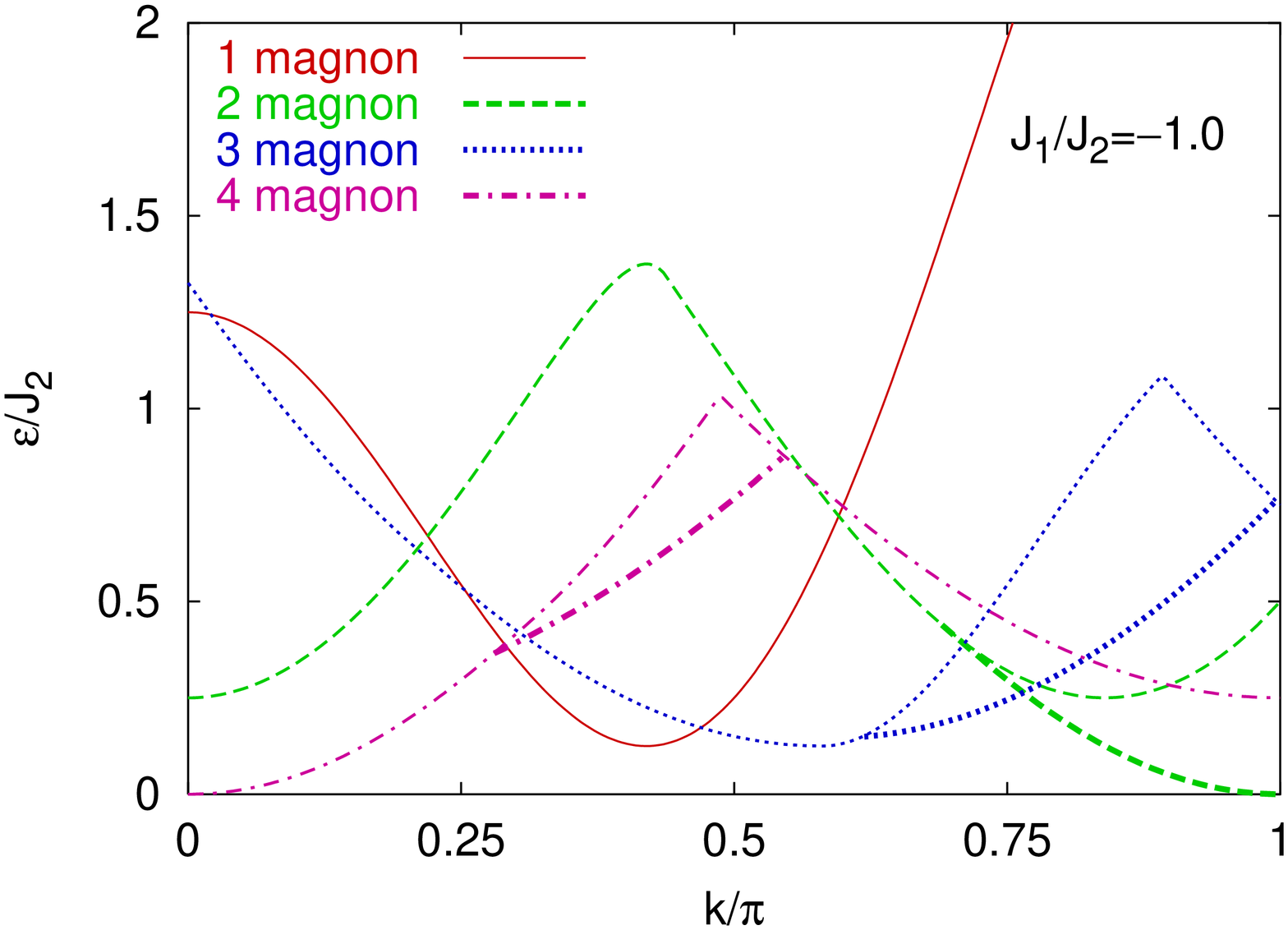,angle=270,width=8.5cm}}
\centerline{\epsfig{file=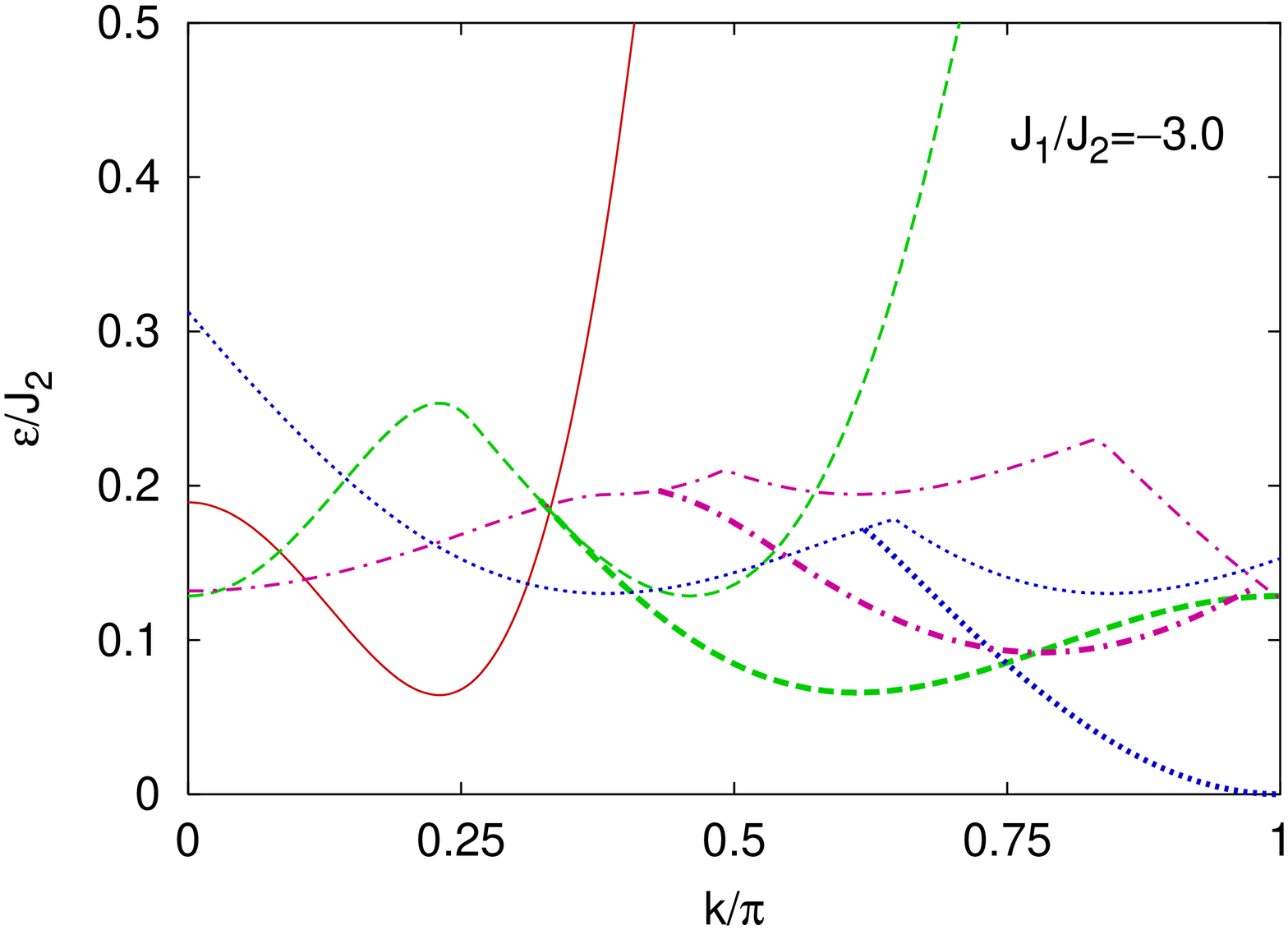,angle=270,width=8.5cm}}
\end{minipage}
\begin{minipage}{8.5cm}
\centerline{\epsfig{file=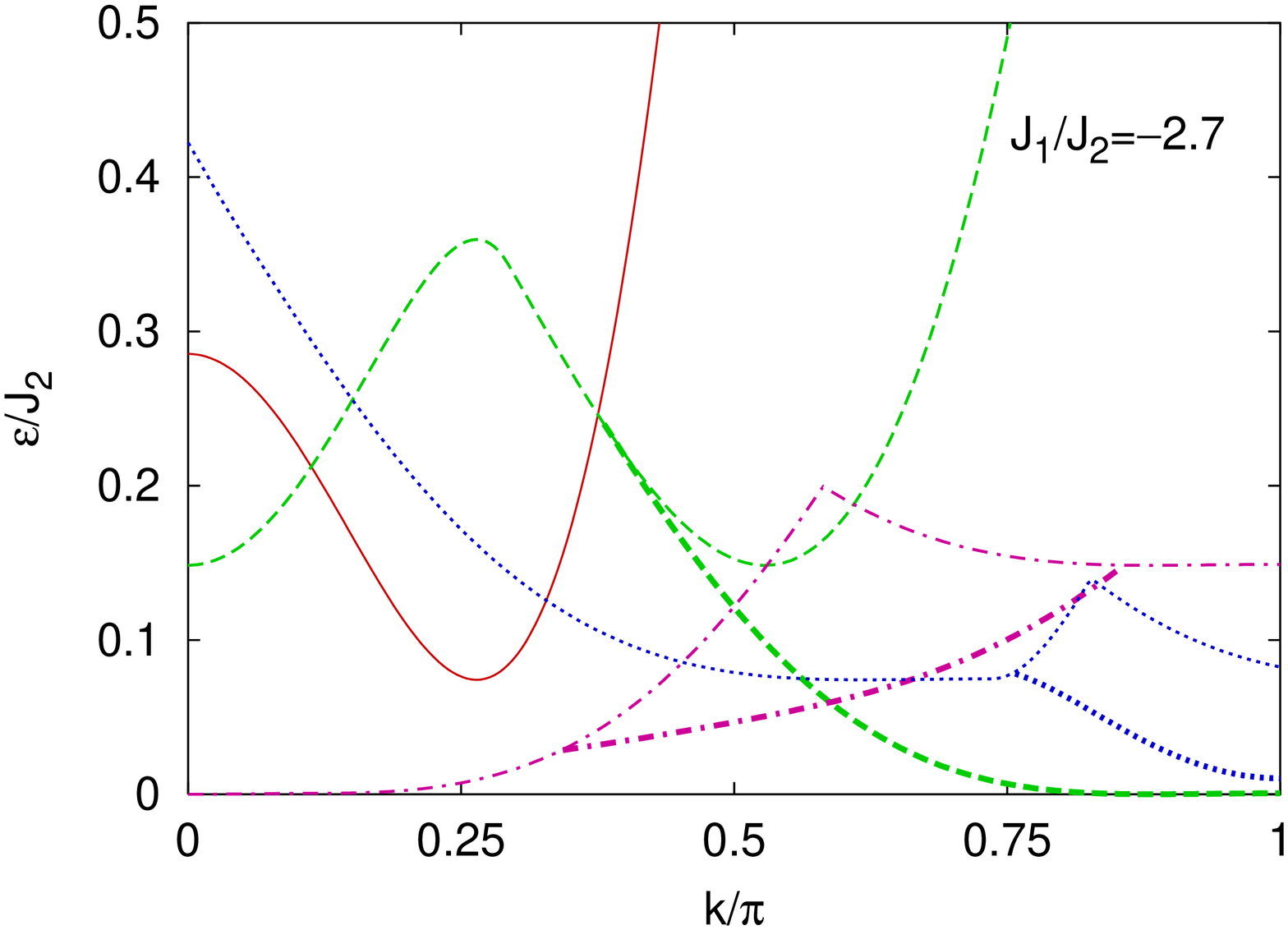,angle=270,width=8.5cm}}
\centerline{\epsfig{file=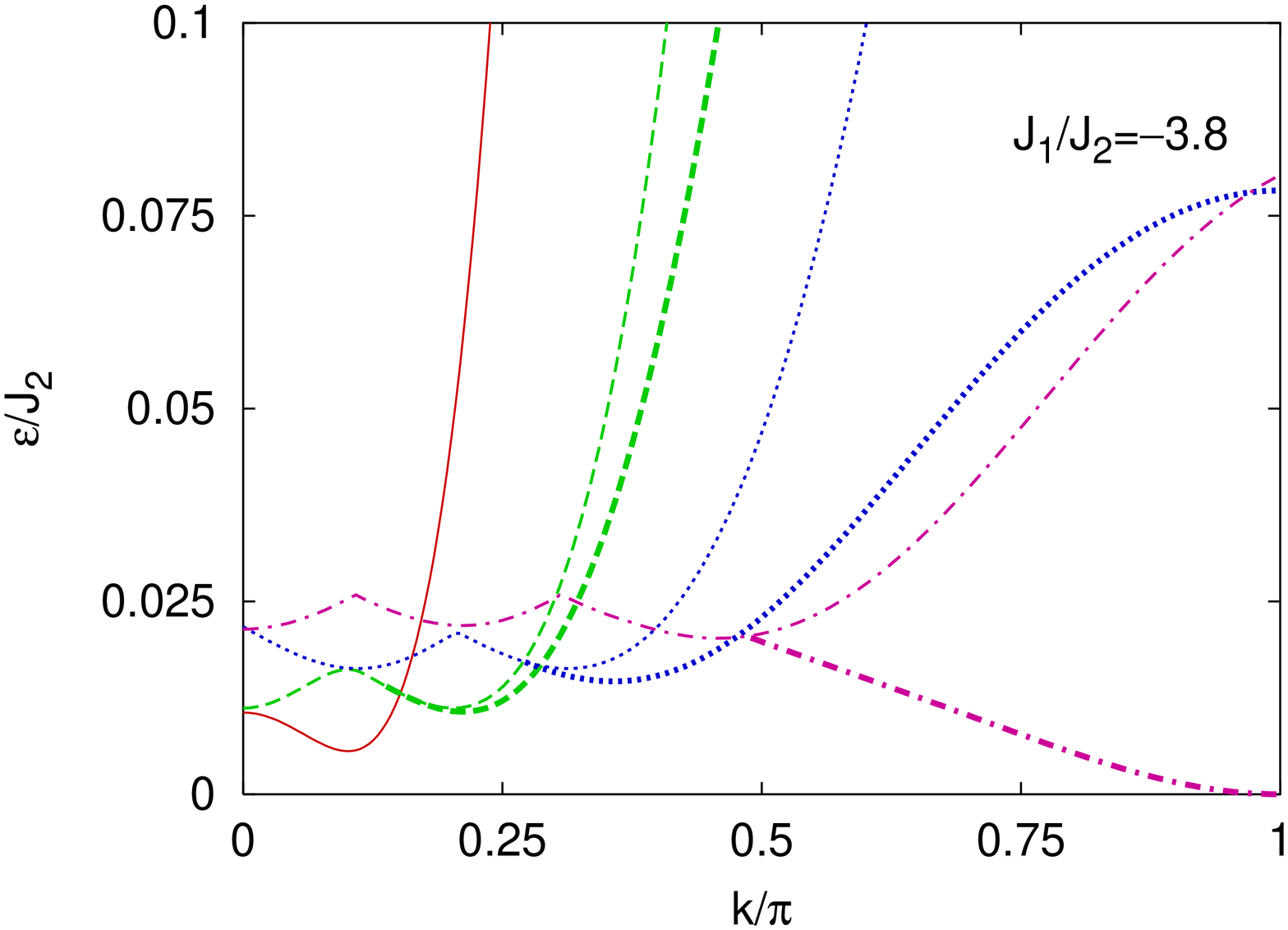,angle=270,width=8.5cm}}
\end{minipage}
\caption{(Color online)
Energy dispersions for the multi-magnon bands at the saturation
field.
Thin lines denote the onset of the scattering continuum,
and thick lines show n-magnon bound states where relevant.
With decreasing $J_1/J_2$, the lowest excitations change
from
commensurate ($k=\pi$) two-magnon bound states,
incommensurate ($k<\pi$) two-magnon bound states,
commensurate three-magnon bound states,
to commensurate four-magnon bound states.
}
\label{dispfig}
\end{figure*}

The four panels of Fig.~\ref{dispfig} show energy dispersions of
multi-magnon excitations at $J_1/J_2=-1.0,$ $-2.7$, $-3.0$, and $-3.8$.
For each value of $J_1/J_2$ the magnetic field is set equal to
the saturation field where
the lowest-lying excited state is gapless.
The dispersion of bound $n=2,3,4$ magnons are obtained
by diagonalizing the Hamiltonian in the finite basis,
as we described above.
Our numerical calculation reproduces the known results
for the two-magnon sector:\cite{chubukov,dmitriev,kuzian}
the lowest bound two-magnon excitation
has momentum $k=\pi$ for $-2.67\lesssim J_1/J_2<0$ and incommensurate
momentum $k<\pi$ for $-4<J_1/J_2\lesssim-2.67$.
Figure \ref{dispfig} also shows the lower edges (thin lines) of
the continuum spectra of scattering states made up of $n$ magnons,
which are calculated from the dispersion of a single magnon
(\ref{epsilon_1(k)}) and those of bound states of up to $n-1$ magnons.
For example, energies of 3-magnon scattering states
are obtained by adding those of either three unbound magnons
or of a 2-magnon bound state plus a free magnon, under the
condition that the total momentum is $k$ (mod $2\pi$).

We see in Fig.~\ref{dispfig} that for any $J_1/J_2$ there is always
a region in the $k$ space where bound states lie well below the scattering
continuum.
The character of the lowest-lying bound states signalling the
instability of the fully polarized state changes
with $J_1/J_2$.
Unlike previously thought, the system shows rather different regimes
as a function of $J_1/J_2$:
a two-magnon commensurate ($k=\pi$) instability ($-2.67<J_1/J_2<0$),
a two-magnon incommensurate instability
as predicted in Ref.~\onlinecite{chubukov} ($-2.72<J_1/J_2<-2.67$),
a three-magnon commensurate instability ($-3.52<J_1/J_2<-2.72$),
and a four-magnon commensurate instability
($-3.86<J_1/J_2<-3.52$).\cite{threeeightsix}
As $J_1/J_2$ approaches further towards $-4$, we expect to have
more-than-four-magnon commensurate phases,
but these are outside the scope of our numerics.
We note that, except in the incommensurate regime
($-2.72<J_1/J_2<-2.67$), the bottom of the lowest-lying mode is
at $k=\pi$.
For example, at $J_1/J_2=-1.0$ the lowest energy state in the 2-magnon
sector is the bound state with $k=\pi$.
The 4-magnon scattering states of two such two-magnon bound states
therefore have lowest energy (which vanish when $h=h_s$) at $k=0$.
The fact that we do not have a mode of 4-magnon bound states
near $k=0$ below the continuum indicates that the interaction
between two 2-magnon bound states of $k=\pi$ is repulsive,
and therefore these 2-magnon bound states are stable.
At $J_1=-3J_2$ we have the 3-magnon bound state with $k=\pi$ as
the lowest-energy excitation.
The presence of stable 3- and 4-magnon bound states provides natural
explanation for the $\Delta S^z>2$ jumps found in the DMRG study of
Ref.~\onlinecite{honecker}.

\begin{figure}[t]
\centerline{\epsfig{file=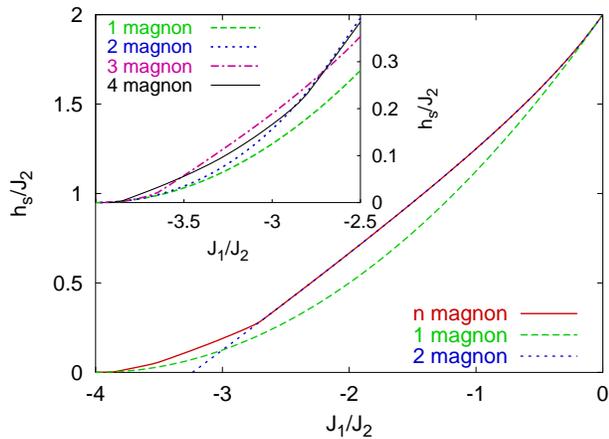,angle=270,width=8.5cm}}
\caption{(Color online)
Saturation field versus $J_1/J_2$.
The solid curve ``$n$ magnon" shows the saturation
field $h_s$ obtained from the numerical $n$-magnon solutions ($n=2,3,4$),
while the curves ``1 magnon" and ``2 magnon" denote $h_s$
calculated from Eqs.\ (\ref{e1mag}) and (\ref{e2mag}), respectively.
The inset shows the region
where our saturation field deviates from the two-magnon solution and
the saturation fields for each multi-magnon excitation.}
\label{hfig}
\end{figure}

Figure \ref{hfig} shows the saturation field $h_s$
determined by the instability from the softening of the lowest excitations.
The saturation field estimated from a single-magnon
instability is always smaller than the true saturation field which
is determined by multi-magnon bound states.
The calculated saturation field is in perfect agreement with
the exact $h_s$ estimated from
Eq.\ (\ref{e2mag}) for $-2.67\lesssim J_1/J_2<0$.
For $-3.86<J_1/J_2<-2.72$ the saturation field is determined from
the instability of 3- or more-magnon bound states.

\begin{figure}[t]
\centerline{\epsfig{file=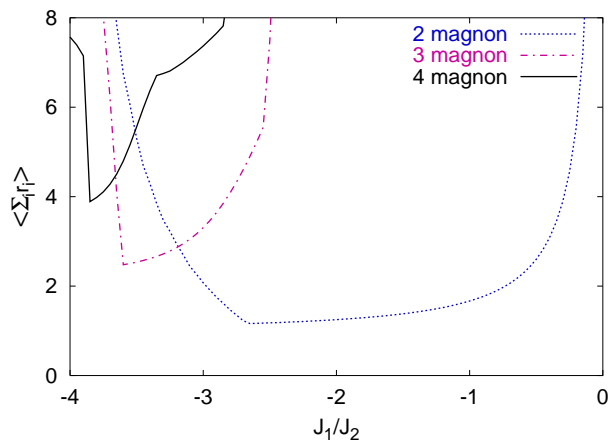,angle=270,width=8.5cm}}
\caption{(Color online) Mean length $\langle\sum_i r_i\rangle$ of the bound
multi-magnon states versus $J_1/J_2$.
The kinks in the curves appear when the total momentum of the
$n$-magnon bound state changes between incommensurate
(where $\langle\sum_i r_i\rangle$ is larger) and commensurate ($k=\pi$).
}\label{rfig}
\end{figure}

In Fig.~\ref{rfig} we show the expectation value
$\langle\sum_{i=1}^{n-1} r_i\rangle$ characterizing the size of the
$n$-magnon bound states, where the average is taken for the
lowest-energy bound state of $n$ magnons for a given
$J_1/J_2$; the minimum value of the average is by definition $n-1$
for the $n$-magnon bound state. We confirm that the magnons are
tightly bound when they are the lowest-lying excitations in the
energy spectra. This justifies our use of basis states with finite
$r_i$ for calculating the low-lying bound states.

We now discuss the implications of the multi-magnon instabilities
that lead to condensation of bound magnons below the saturation
field. We argue that the bound $n$-magnon condensation gives rise to
a phase with multipolar-like quasi-long-range order but without spin
(dipole) ordering in the XY direction. To be specific, let us
consider the case where the $n$-magnon bound states with momentum
$k=\pi$ become gapless as $h\to h_s+0$; $n=2$ for
$-2.67<J_1/J_2<0$, $n=3$ for $-3.52<J_1/J_2<-2.72$, and $n=4$ for
$-3.86<J_1/J_2<-3.52$. Just below the saturation field, the system
can be viewed as a dilute gas of repulsively interacting bosons
which represent the $n$-magnon bound states.\cite{repulsion} Here
the boson creation operator $b_i^\dagger$ is identified with
$(-1)^iS^-_iS^-_{i+1}\cdots S^-_{i+n-1}$ in crude approximation.
These bosons condense to form a TL liquid\cite{repulsion} in which
various correlation functions of the boson fields decay algebraically.
The most slowly decaying two-point correlation function will be the
propagator of the bosons $\langle b_0 b_r^\dagger\rangle$,
which decays as $(-1)^r r^{-1/\eta}$ with\cite{eta} $\eta\to2$
as $h\to h_s-0$.
However, the spin operators which cannot be simply represented with
bosons $b_i$, such as $S_i^-$ and its products
$\Pi_{j=i}^{i+p}S_j^-$ with $p=0,1,\cdots,n-2$, should have only
short-range correlations. This is because exciting one magnon costs
binding energy. Since $n$ down spins form a tightly bound state, we
may use the approximation $\frac12-S_i^z=nb_i^\dagger b_i^{}$. This
allows us to calculate the correlation function of $S_i^z$ from the
density correlation of bosons, which has the asymptotic form
$\langle b_0^\dagger b^{}_0 b_r^\dagger b^{}_r\rangle\sim
\rho^2+Ar^{-\eta}\cos(2\pi \rho r)-\eta(2\pi r)^{-2}$, where $\rho$
is the boson density ($n\rho=\frac12-\langle S^z\rangle$)
and $A$ is a constant. For the case $n=2$, the above
theory indicates that the TL liquid has nematic quasi-long-range
order,\cite{chubukov} which is indeed found by the recent DMRG
calculation at $J_1=-J_2$.\cite{vekua,eta}  The theory also predicts
that the TL liquid with larger $n$ ($n=3,4$) should exist for larger
$|J_1|/J_2$ in the phase diagram of the FM $J_1$-$J_2$ model near
the saturation field. For $n=3$ this is a TL liquid with
antiferro-triatic quasi-long-range order, which is a one-dimensional
analogue of triatic order found in Ref.~\onlinecite{momoi}. Our
numerics suggests that, as $J_1/J_2$ approaches $-4$, instability
from bound states of more magnons appears. We do not know how far
these new phases extend to lower fields.

In summary we have numerically calculated many-magnon bound states
and determined their energy dispersions. We have found that the
fully polarized FM state has instabilities to bose condensation of
these many-magnon bound states, which lead to TL liquids with
multipolar magnetic correlations below the saturation field. It is
our pleasure to acknowledge stimulating discussions with Philippe
Sindzingre and Nic Shannon. This work was in part supported by Japan
Society for Promotion of Science (Grant No.~P06902) and by
Grants-in-Aid for Scientific Research from MEXT (Grant No.~16GS0219
and No.~17071011).


\begin{thebibliography}{99}
\bibitem{threeeightsix} The value of $J_1=-3.86J_2$ for the lower edge
of the 4-magnon region is a lower bound. At $J_1<-3.86J_2$
a loosely bound incommensurate 4-magnon state has
lower energy than the commensurate bound state.
It is quite possible that 5-magnon bound states with $k=\pi$
becomes lowest excitation
at a slightly larger value, $J_1/J_2\approx-3.8$.
\bibitem{repulsion} Assumption of mutual repulsive interaction
is justified by the absence of a metamagnetic transition below
the saturation field and by the $\Delta S=n$ jumps in
the magnetization curve.\cite{honecker,cabra}
For the repulsive interaction the TL parameter $K$ is less than 1.
\bibitem{eta}
The exponent $\eta$ is related to $K$ as $\eta=2K$.
Reference \onlinecite{vekua} obtained $\eta=1.1$ for $J_1=-J_2$ ($n=2$)
and $\langle S^z\rangle=3/8$.
\end{thebibliography}
\end{document}